%% file: ckm2010proc.tex
\documentclass[12pt]{article}
\usepackage{graphicx}
\newcommand{\abs}[1]{\ensuremath{\left|#1\right|}}
\newcommand{\BR} {\ensuremath{\mathrm{ BR}}}
\newcommand{\ks}  {\ensuremath{K_{S}}}
\newcommand{\kl}  {\ensuremath{K_{L}}}
\newcommand{\kzero} {\ensuremath{K^{0}}}
\newcommand{\kpm} {\ensuremath{K^{\pm}}}
\newcommand{\vud} {\ensuremath{V_\mathrm{ud}}}
\newcommand{\vus} {\ensuremath{V_\mathrm{us}}}
\newcommand{\vub} {\ensuremath{V_\mathrm{ub}}}
\newcommand{\fVus}{\ensuremath{f_+(0)\,V_{us}}}
\newcommand{\fzero}  {\ensuremath{f_+(0)}}
\newcommand{\fzerokpi}  {\ensuremath{f_+^{K^0\pi^-}(0)}}
\newcommand{\lv}       {\ensuremath{\lambda_+}}
\newcommand{\ls}       {\ensuremath{\lambda_0}}
\newcommand{\kltre}  {\ensuremath{K_{l3}}}
\newcommand{\ksetre}  {\ensuremath{K_{Se3}}}
\newcommand{\taus}  {\ensuremath{ \tau_S}}
\newcommand{\taul}  {\ensuremath{\tau_L}}
\newcommand{\taupm}  {\ensuremath{\tau_{\pm}}}
\newcommand{\mudue}  {\ensuremath{\mu^{\pm}\nu}}
\newcommand{\pipm} {\ensuremath{\pi^{\pm}}}
\newcommand{\pippim}  {\ensuremath{\pi^+\pi^-}}
\def\ff{$\phi-$factory}  \def\DAF{DA\-$\Phi$\-NE}

\def\ifm#1{\relax\ifmmode#1\else$#1$\fi}
  
\def\pt#1,#2,{\ifm{#1\x10^{#2}}}
\def\L{\ifm{{\cal L}}}
\def\dt{\ifm{{\rm d}\,t}}
\def\epm{\ifm{e^+e^-}}
\def\lnf{Laboratori Nazionali di Frascati dell'Istituto Nazionale di Fisica Nucleare,\\
00044 Frascati (RM), ITALY}

\def\Title#1{\begin{center} {\Large #1 } \end{center}}
\def\Author#1{\begin{center}{ \sc #1} \end{center}}
\def\Address#1{\begin{center}{ \it #1} \end{center}}

\newenvironment{Abstract}{\begin{quotation}  }{\end{quotation}}
\newenvironment{Presented}{\begin{quotation} 
\bigskip 
      \begin{center}\begin{large}}{\end{large}\end{center} \end{quotation}}

\input econfmacros.tex

\begin{document}
\begin{titlepage}

\vfill
\Title{Determination of \vus\ at the KLOE experiment: present results and future perspectives}
\vfill
\Author{ Erika De Lucia for the KLOE and KLOE-2 Collaborations}
\Address{\lnf}
\vfill
\begin{Abstract}
Precise measurements of semileptonic kaon decay rates at KLOE
provide the measurement of the CKM mixing matrix  element \vus\ 
and information about lepton universality. 
Leptonic kaon decays provide 
an independent measurement of $\abs{\vus}^2/\abs{\vud}^2$, through  
the ratio  $\Gamma(K\to\mu \nu)/\Gamma(\pi\to\mu \nu)$.
These measurements, together with the
result of $|\vud|$ from nuclear $\beta$ transitions,
provide the most precise test of CKM unitarity,
allowing the universality of lepton and quark weak couplings to be tested. 
After the completion of the KLOE data taking, the proposal
of a new run with an upgraded KLOE detector, KLOE-2, at an upgraded \DAF\ machine has been accepted by INFN and it is now starting.  Present results from KLOE and future perspectives from KLOE-2 are reported.
\end{Abstract}
\vfill
\begin{Presented}
Proceedings of CKM2010, the 6th International Workshop on the CKM Unitarity Triangle, University of Warwick, UK, 6-10 September 2010
\end{Presented}
\vfill
\end{titlepage}
\def\thefootnote{\fnsymbol{footnote}}
\setcounter{footnote}{0}

\section{Introduction}
The KLOE experiment~\cite{bib:rivista} collected an integrated luminosity
$\int\!\L\dt\,$ $\sim$2.5fb$^{-1}$ at the Frascati \ff\  \DAF , an \epm\ collider operated at the
energy of 1020 MeV, the $\phi$-meson mass. 
With its general purpose detector, consisting of a large cylindrical drift chamber surrounded by a
lead-scintillating fiber electromagnetic calorimeter entirely
immersed in an axial magnetic field,
KLOE produced the most comprehensive set of results on kaon physics
from a single experiment using the unique availability of pure \ks\ ,\kl\
and \kpm\ beams at a \ff\. After the completion of the KLOE data taking, the proposal of a new run with an upgraded KLOE detector, KLOE-2~\cite{kloe2eoi}, at an upgraded \DAF\ machine has been accepted and it is now starting.

\par
An overview of KLOE results for \kl, \ks\ and \kpm\ used to extract 
\vus\ is presented (sec.~\ref{sec:obs})
 together with the future perspectives within the KLOE-2 project (sec.~\ref{sec:future}).
\section{\vus\ from kaon decays: unitarity and universality}
\label{sec:obs}
The most precise test of CKM unitarity is given by the constraint on its first row $|\vud|^2 + |\vus|^2 + |\vub|^2=1$
with $|\vud|$ measured from superallowed $0^+ \rightarrow 0^+$ nuclear $\beta$ transitions, 
$|\vus|$ from semileptonic kaon decays and $|\vub|^2$ being negligible.
The kaon semileptonic decay rate is given by:  
\begin{equation}
\Gamma(\kltre) = \frac{C_K^2 G_F^2 M_K^5}{192 \pi^3} S_{EW} |\vus|^2 |\fzero|^2 
I_{K,l}(\lambda)(1+2\Delta_K^{SU(2)}+2\Delta_{K,l}^{EM})
\label{eq:gammakl3}
\end{equation}
where $K = \kzero, \kpm$, $l = e, \mu$ and $C_K$ is a Clebsch-Gordan coefficient, equal to $1/2$ and $1$ for \kpm\ 
and \kzero, respectively. The decay width $\Gamma(\kltre)$ is experimentally determined 
by measuring the kaon lifetime and the semileptonic \BR s, inclusive of radiation. The theoretical inputs are: the universal 
short-distance electroweak correction $S_{EW} = 1.0232$, 
the $SU(2)$-breaking $\Delta_K^{SU(2)}$ and the long-distance electromagnetic
corrections $\Delta_{K,l}^{EM}$,
 and the form factor $\fzero\equiv\fzerokpi$ 
evaluated at zero momentum transfer. 
 The form factor dependence on the momentum transfer can be described by one or more slope parameters $\lambda$, 
measured from the decay spectra, and enters in the phase space integral
$I_{K,l}(\lambda)$.
\begin{table}[!h!b!t]
\begin{center}
\begin{tabular}{ll}\hline\hline
\multicolumn{2}{c}{Branching ratios}\\
\hline 
{\em $K_L \to \pi e \nu$}   & 0.4008 $\pm$ 0.0015 \\ 
 {\em $K_L \to \pi\mu\nu$}   & 0.2699 $\pm$ 0.0014 \\ 
 {\em $K_S \to \pi^{+}\pi^{-}$}   & 0.60196 $\pm$ 0.00051 \\ 
 {\em $K_S \to \pi^{0}\pi^{0}$}   & 0.30687 $\pm$ 0.00051 \\ 
 {\em $K_S \to \pi e \nu$}   & (7.05 $\pm$ 0.09)$\times$10$^{-4}$ \\ 
 {\em $K^{+} \to \mu^{+}\nu(\gamma)$}   & 0.6366 $\pm$ 0.0017 \\ 
 {\em $K^{+} \to \pi^{+}\pi^{0}(\gamma)$}   & 0.2067 $\pm$ 0.0012 \\ 
 {\em $K^{+} \to \pi^{0}e^{+}\nu(\gamma)$}   & 0.04972 $\pm$ 0.00053 \\ 
 {\em $K^{+} \to \pi^{0}\mu^{+}\nu(\gamma)$} & 0.03237 $\pm$ 0.00039 \\ 
\hline\hline
\end{tabular}
\kern1cm
\begin{tabular}{ll}\hline\hline
\multicolumn{2}{c}{Lifetimes and Form factors}\\
\multicolumn{2}{c}{     (dispersive approach)}\\
\hline
$\taul$   & 50.84 $\pm$ 0.23   \\ 
$\taupm$    &  12.347 $\pm$ 0.030 \\ 
\lv & 25.7(0.6)$\times10^{-3}$ \\ 
\ls &  14.0(2.1)$\times10^{-3}$ \\ 
\hline\hline\end{tabular}
\caption{Summary of KLOE results useful for \vus\ measurement~\cite{bib:rivista}.}
\label{TAB:res}
\end{center}
\end{table}
\par 
All the relevant inputs to extract \vus\ from \kltre\ decay rates 
have been measured at KLOE~\cite{bib:rivista}: branching ratios (\BR s), lifetimes and form factors (Table \ref{TAB:res}).
Complementary to \kltre\ decays, the measurement of $\BR(\kpm \rightarrow \mudue)$ allowed 
us to extract $\vus/\vud$ and the result of $\BR(K^{+} \to \pi^{+}\pi^{0}(\gamma))$ improved the accuracy. 
Recently KLOE has also measured the \ks\ lifetime from the fit to the proper time distribution obtained with a sample of $\sim$20 million $\ks \to \pippim$ decays. The final result, presently the most precise, is $\taus = 89.562\pm9.029_{\rm stat}\pm0.043_{\rm syst}$ ps \cite{kslifetime}.
%
To  extract $\vus\fzero$ we use eq.\ref{eq:gammakl3} together with the $SU(2)$-breaking and 
long distance $EM$ corrections to the full inclusive decay rate~\cite{su2breaking}.
The measured values of $\vus\fzero$ are~\cite{allvus}: 0.2155(7) for $K_Le3$, 0.2167(9)
for $K_L\mu3$, 0.2153(14) for $K_Se3$, 0.2152(13) for $K^{\pm}e3$, and
0.2132(15) for $K^{\pm}\mu3$ decays. 
Their average is $\vus\fzero$ = 0.2157(6) ($\chi^2/ndf =7.0/4$,Prob=13\%),
with 0.28\% accuracy to be compared with the 0.23\% of the world average $\vus\fzero$ = 0.2166(5)~\cite{bib:flavianet}.
Defining $r_{\mu e}=|\fVus|_{\mu3}^2/|\fVus|_{e3}^2=g_{\mu}^2/ g_{e}^2$,
with $g_{\ell}$ the coupling strength at the $W \to \ell \nu$ vertex,
lepton universality can be tested comparing the measured value with
its Standard Model (SM) prediction $r_{\mu e}^{SM}=1$. 
We obtain $r_{\mu e} =1.000(8)$, averaging between charged and neutral
modes, to be compared with
$(r_{\mu e})_{\pi} =1.0042(33)$ from leptonic pion decays, and 
$(r_{\mu e})_{\tau} =1.0005(41)$ from leptonic $\tau$
decays~\cite{unilep_pi_tau}.
%
Using $\vus\fzero$ from \kltre\ decays and 
$\fzero = 0.964(5)$~\cite{UKQCD}, 
we get $\vus = 0.2237(13)$. 
Furthermore $\vus/\vud$ can be measured using the radiative inclusive decay 
rates of $\kpm \rightarrow \mudue (\gamma)$ and $\pipm \rightarrow \mudue (\gamma)$, combined 
with a lattice calculation of $f_K/f_{\pi}$.
Using $\BR(\kpm \rightarrow \mudue) = 0.6366(17)$ from KLOE~\cite{kmu2}
 and $f_K/f_\pi = 1.189(7)$~\cite{HP/UKQCD},
we get $\vus/\vud = 0.2323(15)$. 
Combining this result with \vus\ from \kltre\ decays and $\vud =
0.97418(26)$~\cite{vudyetold}, CKM unitarity has been verified to the level of $1-\vud^2 + \vus^2+\vub^2=4(7)\times10^{-4}$. 
We then obtained $G_{\rm CKM}=G_{\rm F} (\vud^2 + \vus^2+\vub^2)^{1/2}=1.16614(40)\times
10^{-5}$ GeV$^{-2}$, is in perfect agreement with the measurement from the muon
lifetime $G_{\rm F}=1.166371(6)\times 10^{-5}$ GeV$^{-2}$. This result significantly improves the accuracy obtained with evaluations from tau-lepton decays and electroweak precision tests. 
\section{The KLOE-2 project}
\label{sec:future}
The KLOE-2 project aims at improving the successful and fruitful results
achieved by the KLOE Collaboration in Kaon and Hadron Physics and extending the
physics program~\cite{kloe2paper} to: $\gamma \gamma$-physics from \epm\ $\to$ \epm\ $\gamma^* \gamma^* \to$ \epm\ +$X$ and search for particles from hidden sectors that might explain dark matter. The project will exploit the new interaction scheme implemented and tested on the Frascati \DAF\ collider with the SIDDHARTA experiment in 2009 \cite{crabwaist} with larger beam crossing angle and crab-waist sextupoles. This allowed a luminosity increase of factor of $\sim$3 to be reached with a peak luminosity $\L$=4.5x$10^{32}$ cm$^{-2}$s$^{-1}$ and an integrated luminosity $\int\!\L\dt\,$ $\sim$1pb$^{-1}$/h. With this new configuration $\int\!\L\dt\,$ $\sim$5fb$^{-1}$/y can be delivered.
\par After a first phase with the installation of the low-energy $e^+e^-$
(LET)~\cite{bib:tdr_gg,bib:let} and  high-energy $e^+e^-$ (HET)~\cite{bib:tdr_gg,bib:let_het} tagging
systems for the identification and study of $\gamma\gamma$ events,
the detector will be upgraded with the insertion of an
Inner Tracker (IT)~\cite{bib:tdr_it,bib:nim_it}, between the beam pipe and the Drift 
Chamber (DC) inner wall, and with crystal calorimeters (CCALT)~\cite{bib:ccalt}, covering the low $\theta$ region, and two new tile calorimeters (QCALT)~\cite{bib:qcalt} instrumenting the \DAF\ focusing system.
\par Using KLOE present data set together with the 5 fb$^{-1}$ KLOE-2/{\rm step0} foreseen statistics, we can improve the accuracy with respect to present world average~\cite{bib:flavianet} on the measurement of \kl\ , \kpm\ lifetimes and \ksetre\ branching ratio, presently the main contributors to \fVus\ uncertainties. Statistical uncertainties on \BR s and lifetimes have been obtained scaling present statistics to 7.5 fb$^{-1}$ total integrated luminosity and a conservative estimate of systematic errors has been obtained based on KLOE published analyses, without improvements from detector upgrades. Systematic errors in KLOE are partially statistical in nature, efficiencies are measured with data control samples, then also these contributions to the total uncertainty decrease with statistics.
\par The accuracy on the measurement of \taul\ from the fit to the proper time distribution of $K_L \to 3\pi^0$ decays is expected to be reduced to 0.27\% and furthermore below 0.2\% with the  QCALT insertion, improving photon reconstruction and control of systematic effects.  With 7.5 fb$^{-1}$ total integrated luminosity a 0.1\% accuracy on the \taupm\ measurement is expected to be reached and a factor of $\sim$2 improvement with the IT detection of \kpm\ tracks closer to the interaction point, improving the accuracy of the decay length technique.
\par The branching ratio of \ksetre\ decays is expected to be measured with 0.6\% accuracy and further improved to 0.3\% with the IT.
As a matter of fact the measurement of \kltre\ decay rates will strongly benefit from the insertion of the IT detector:
this upgrade will increase the acceptance for decays close to the interaction point with low momentum tracks and improve the resolution on their
track and vertex parameters. 
\par In conclusion, a significant reduction of the present experimental uncertainty on \vus\fzero\ is expected: the present 0.23\% fractional uncertainty on \vus\fzero\ can be reduced to 0.14\%, using KLOE present data set together with the KLOE-2/{\rm step0} statistics. Detector upgrades have not been considered in this evaluation. This, together with more precise measurements of \fzero\ and \vud\,
would allow us to reach the level of precision of a few 10$^{-4}$ in the test on the unitarity relation thus improving the potential to investigate new physics within SM extensions with gauge universality breaking. 


\end{document}

%% file: econfmacros.tex
\def\beq{\begin{equation}}
\def\eeq#1{\label{#1}\end{equation}}
\def\eeqn{\end{equation}}

\def\beqa{\begin{eqnarray}}
\def\eeqa#1{\label{#1}\end{eqnarray}}
\def\eeqan{\end{eqnarray}}

\let\bar=\overbar

\def\L{{\cal L}}

\def\Dslash{\not{\hbox{\kern-4pt $D$}}}
\def\dslash{\not{\hbox{\kern-2pt $\del$}}}

\def\BR{\mbox{\rm BR}}

\def\msb{{\bar{\ssstyle M \kern -1pt S}}}

